\documentclass[prd,twocolumn,nofootinbib,amsmath,amsfonts,amssymb,showpacs,preprintnumbers]{revtex4-1} %% Journal version

\usepackage[usenames]{color}
\usepackage{epsfig,graphicx,bm,color}

\usepackage{natbib}
\newcommand{\citeeq}[1]{Eq.~(\ref{#1})}

\newcommand{\citeeqss}[2]{Eqs.~(\ref{#1})~and~(\ref{#2})}

\newcommand{\citeeqp}[1]{Eq.~\ref{#1}}
\newcommand{\citesec}[1]{Sec.~\ref{#1}}

\newcommand{\citetab}[1]{Table~\ref{#1}}
\newcommand{\citefig}[1]{Fig.~\ref{#1}}

\newcommand{\eg}{{\it e.g.}}
\newcommand{\ie}{{\it i.e.}}

\newcommand{\ben}{\begin{eqnarray}}
\newcommand{\een}{\end{eqnarray}}
\newcommand{\be}{\begin{equation}}
\newcommand{\ee}{\end{equation}}
\newcommand{\nn}{\nonumber}

\newcommand{\msun}{\mbox{$M_{\odot}$}}

\newcommand{\mchi}{\mbox{$m_{\chi}$}}
\newcommand{\sigv}{\mbox{$\langle \sigma v \rangle $}}
\newcommand{\rsun}{\mbox{$r_\odot$}}

\graphicspath{{./NewFigs/}}
\begin{document}

\title{Sunyaev-Zel'dovich effects from annihilating dark matter in the 
  Milky Way:\\ smooth halo, subhalos and intermediate-mass black holes}

\author{Julien Lavalle}
\affiliation{Dipartimento di Fisica Teorica, Universit\`a di Torino \& INFN,
  via Giuria 1, 10125 Torino --- Italy}
\email{lavalle@to.infn.it}
\email{current address: IFT-Madrid, Spain}

\date{30 June 2010}

\begin{abstract}
We study the Sunyaev-Zel'dovich effect potentially generated by 
relativistic electrons injected from dark matter annihilation
or decay in the Galaxy, and check whether it could be observed by Planck or 
the Atacama Large Millimeter Array (ALMA), or even imprint the current CMB data 
as, \eg, the specific fluctuation excess claimed from an recent reanalysis of 
the WMAP-5 data. We focus on high-latitude regions to avoid contamination of the
Galactic astrophysical electron foreground, and consider the annihilation or 
decay coming from the smooth dark matter halo as well as from subhalos, further 
extending our analysis to a generic modeling of spikes arising around 
intermediate-mass black holes. We show that all these dark Galactic components 
are unlikely to produce any observable Sunyaev-Zel'dovich effect. For a 
self-annihilating dark matter particle of 10 GeV with canonical properties, the 
largest optical depth we find is $\tau_e \lesssim 10^{-7}$ for massive isolated 
subhalos hosting intermediate-mass black holes. We conclude that dark matter 
annihilation or decay on the Galactic scale cannot lead to significant 
Sunyaev-Zel'dovich distortions of the CMB spectrum.
\end{abstract}

\pacs{95.35.+d, 97.60.Lf, 96.50.S-}

\maketitle

\preprint{DFTT-11/2010}

\section{Introduction}
\label{sec:intro}

The Sunyaev-Zel'dovich (SZ) effect stands for the distortion of the cosmic 
microwave background (CMB) spectrum by the scattering of thermal or 
relativistic electrons\footnotemark~\cite{1972CoASP...4..173S}. In this paper, 
we aim at studying the SZ effect as a possible signature complementary to 
other indirect probes of dark matter annihilation on the Galactic scale 
(\eg~\cite{1978ApJ...223.1015G,1978ApJ...223.1032S,1984PhRvL..53..624S,1985PhRvL..55..257S,1986PhRvL..56..263S,1986PhRvD..34.1921T}), 
even though it was recently shown to be hardly observable for galaxy clusters
\cite{2009JCAP...10..013Y,2010JCAP...02..005L}. This study is partly motivated
by the recent hint for $\sim 10^\circ$ angular scale SZ signals in the 
WMAP 5-year data away from the Galactic plane by~\cite{2010arXiv1002.4872J}, 
potentially of Galactic origin, which might be explained by a column density
of electrons of $\sim 10^{22}\,{\rm cm^{-2}}$, \ie~an optical depth as large 
as $\sim 10^{-3}$.

\footnotetext{In the following, the term {\em electron} denotes electron or 
positron indifferently, unless specified otherwise.}

A simplistic comparison of the physical scales relevant to the calculation 
already provides some interesting information, since the SZ signal is roughly 
independent of the target distance (except for angular resolution effects). In 
galaxy clusters, the main contribution to the SZ signal, already difficult to 
observe to high precision, comes from a typical {\em thermal} electron density 
$n_e^{\rm th} \sim 10^{-2}\,{\rm cm^{-3}}$ integrated along a line-of-sight 
spatial scale $l\sim 500\,{\rm kpc}$. The SZ effect amplitude is roughly
set by the electron optical depth $\tau_e \sim \sigma_{\rm T}\, n_e\, l$, where 
$\sigma_{\rm T}$ is the Thomson cross section. With the previous numbers
we readily get $\tau_e \sim 10^{-3}$ for thermal electrons in clusters, 
consistent with most predictions (\eg~\cite{1999PhR...310...97B}), which
provides a reference value for detectability. In the Milky Way, the typical 
{\em relativistic} electron density measured at the Earth, which constitutes a 
sound {\em local} upper bound to the yield potentially originating from dark 
matter annihilation at the GeV-TeV energy scale, is $\sim 10^{-11}-10^{-12}\,
{\rm cm^{-3}}$ around 1 GeV (see \eg~\cite{2000PhLB..484...10A}); this density 
can be associated with a typical spatial scale of a few tens of kpc for the 
bulk of usual dark matter density profiles. This translates into 
$\tau_e \lesssim 10^{-12}$ for relativistic Galactic electrons of energy
$> 1$ GeV. Although the lower energy part of the electron spectrum should play 
an important role (see the discussion in \citesec{subsec:bg}), we could at 
zeroth order bet for no significant effect caused by dark matter annihilation 
or decay products at this stage.

Nevertheless, dark matter can collapse on scales as small as its free-streaming 
length at the twilight of the radiation era in the early universe, which is 
much smaller than the size of the Galactic halo for generic weakly interacting 
massive particles (WIMPs) (\eg~\cite{2009NJPh...11j5027B}). Such dark lumps are 
called subhalos and are commonly observed in N-body simulations of structure 
formation (\eg~\cite{2005Natur.433..389D}), though on scales larger than a few
tens of pc due to numerical limits. This clustering implies a significant degree
of inhomogeneity in the Galactic dark matter distribution which may lead to a 
global increase of the annihilation rate~\cite{1993ApJ...411..439S}. 
This may consequently increase the electron density injected by dark matter 
annihilation over the entire Galaxy, and therefore along a given line of sight 
--- in contrast, the effect is expected much less important for decaying dark 
matter for which the decay rate scales only linearly with its density, so that 
the global injection rate is fixed by the total mass of the Galaxy. Indeed, 
although it was shown that subhalos could not drastically enhance the local 
dark-matter-induced electrons~\cite{2008A&A...479..427L}, the SZ effect is a 
cumulative effect along a line of sight which may therefore be more affected 
by subhalos, as it is the case for the complementary gamma-ray 
signals~\cite{1999PhRvD..59d3506B}. Moreover, in addition to this class of 
inhomogeneities, other putative Galactic compact objects like 
intermediate-mass black holes (IMBHs) could raise spikes of dark matter 
\cite{2005PhRvL..95a1301Z} and might also be able to seed SZ features in the 
CMB spectrum. It is not as straightforward as above to estimate the 
contributions of these different components, and it is consequently interesting 
to clarify this issue by means of explicit calculations, which we perform below.

The outline is the following. In \citesec{sec:tau}, we first focus on the 
basic physical parameter that sizes the amplitude of the CMB spectral 
distortion, \ie~the optical depth. In \citesec{sec:dm}, we calculate the 
optical depths for the smooth dark matter halo, and for both the 
resolved and unresolved subhalos. In \citesec{sec:imbhs}, we briefly
discuss the IMBH case, before concluding in \citesec{sec:concl}.

\section{Sizing the SZ amplitude: the electron optical depth}
\label{sec:tau}

Dark matter annihilation in the GeV-TeV energy range leads to the injection of 
relativistic electrons. Interestingly enough, the SZ signal generated by 
these electrons should not suffer too much from the Galactic foreground if 
detected at sufficiently high latitude, since the astrophysical sources of 
electrons are expected to be concentrated in the Galactic disk. Moreover, no 
significant additional thermal SZ is expected to shield the potential dark 
matter contribution, in contrast to the galaxy cluster case.

Two different formalisms were developed to calculate the SZ effect, one based on
radiative transfer (\eg\ \cite{1979ApJ...232..348W,1995ApJ...445...33R,1999PhR...310...97B,2000A&A...360..417E}), 
offering a nice analytical framework for thermal or relativistic electrons as 
long as the Thomson approximation is valid, and another relying on the 
covariant Boltzmann equation for which relativistic corrections to the thermal 
case were often obtained by means of Taylor expansion methods 
(\eg\ \cite{1998ApJ...499....1C,1998ApJ...502....7I}). In fact, these two 
formalisms were recently shown to be equivalent in the Thomson regime 
\cite{2009PhRvD..79h3505B}, in which analyticity is therefore preserved 
\cite{2009PhRvD..79h3505B,2009PhRvD..79h3005N}. In the following, we use the 
formalism presented in Ref.~\cite{2010JCAP...02..005L}, to which we refer the 
reader for more details.

One of the most important physical parameters entering the SZ prediction is the
so-called {\em optical depth} $\tau_e$ characterizing the electrons responsible 
for the spectral distortion of the CMB. Averaging it over the angular resolution
$\delta\Omega$ of the detector, we have
\ben
\langle \tau_e \rangle_{\rm res} &=& 
\frac{\sigma_{\rm T}}{\delta\Omega(\mu_{\rm res})}
\int_{\delta\Omega(\mu_{\rm res})} d\Omega \int dl \, n_e(\vec{x})\;,
\label{eq:def_tau_av}
\een
where $dl$ denotes the line of sight, $\sigma_{\rm T}$ is the Thomson cross 
section, $n_e$ is the electron density in the target, $\mu_{\rm res}\equiv 
\cos(\theta_{\rm res})$ features the angular resolution $\theta_{\rm res}$.
The typical optical depth leading to observable thermal SZ is 
$\tau_e \gtrsim 10^{-3}$ in galaxy clusters~\cite{1999PhR...310...97B}, which 
provides us with a benchmark value useful for further comparisons, while
current and coming experiments can reach micro-Kelvin temperature
fluctuations, or equivalently $\tau_e \sim 10^{-4}$
(\eg~\cite{2006astro.ph..4069T,alma_paper,2009IEEEP..97.1463W}).

In the following, we compute the electron density $n_e$ expected for the 
different dark matter components introduced above.

\section{SZ from the annihilation products of the dark matter halo and subhalos}
\label{sec:dm}

Electrons potentially injected in the Milky Way from dark matter 
annihilation or decay (say on the GeV-TeV energy scale) are expected to diffuse 
on small-scale moving magnetic turbulences and lose their energy through Compton
interactions with the interstellar radiation fields (CMB is one of them) and the
magnetic field, and through Coulomb interactions with the interstellar gas. 
For regions distant by more than a few kpc to the Galactic plane and almost
devoid of interstellar gas and magnetic field, the main target for energy
loss is the CMB. Nevertheless, independent of the peculiar regime, the 
transport equation that describes the evolution of the electron phase-space 
density after injection has to include all important processes --- the 
associated general mathematical formalism is well established 
(\eg~\cite{1964ocr..book.....G,berezinsky_book_90}).

\subsection{Galactic foreground}
\label{subsec:bg}

There exist many astrophysical sources of high-energy electrons in the 
Galactic disk, like supernova remnants or pulsars (see 
\eg~\cite{2010arXiv1002.1910D} for a recent analysis of the local flux),
which justifies to preferentially look for dark-matter-induced SZ signals at 
high Galactic latitude. The SZ foreground due to this specific population can
be grossly assessed by slightly refining the argument discussed in the 
Introduction. The Fermi experiment has recently measured the flux of electrons 
in the energy range 10-1000 GeV \cite{2009PhRvL.102r1101A,2009arXiv0912.3611P}, 
which amounts to $\phi_e(E) \approx 2\times 10^{-5}(E/10\,{\rm GeV})^{-3}\,
{\rm cm^{-2}s^{-1}sr^{-1}GeV^{-1}}$. This flux translates into a density 
of $dn_e^{\rm astro}(E)/dE \approx 8\times 10^{-15} (E/10\,{\rm GeV})^{-3} \,
{\rm cm^{-3}GeV^{-1}}$. A naive integration of this power law down to 1 MeV
provides a reference value of 
$n_e^{\rm astro}(>1\,{\rm MeV}) \approx 4\times 10^{-6}\, {\rm cm^{-3}}$. Assuming
that this density is constant up to $L\sim 5$ kpc in the direction perpendicular
to the Galactic plane, which roughly corresponds to the vertical extent of the 
cosmic-ray confinement region, and vanishes beyond, we can derive an approximate
optical depth of
\ben
\tau_{e,{\rm rel}}^{\rm astro}\approx \sigma_{\rm T} \, n_e^{\rm astro}\,L \approx 
4\times 10^{-8} \;.
\label{eq:tau_bg}
\een
This gives an indication about the high-latitude contribution to the SZ of 
astrophysical relativistic electrons located close to the disk, which the dark 
matter yield in the same region cannot exceed too much without exceeding the 
local electron flux in the meantime. This is likely an overestimate since the 
astrophysical electron density is expected to decrease quite fast away from the 
disk, where most of the sources are located.

To complete the astrophysical foreground picture, we need to account for the 
thermal electron contribution. We take advantage of the well-known NE2001 model 
designed in Ref.~\cite{2002astro.ph..7156C} from pulsar dispersion measures, 
from which it is quite easy to numerically perform the line-of-sight integral 
perpendicular to the Galactic plane, from the Earth location. Taking the thin 
disk and thick disk components of this model, we find that 
\ben
\tau_{e,{\rm th}}^{\rm astro}\approx 10^{-4}\;.
\label{eq:tau_bg_th}
\een
Thus, the local thermal astrophysical foreground is likely the dominant one, 
with a rather large amplitude. Such a value is actually not that surprising 
since it was already emphasized in Ref.~\cite{2003MNRAS.345.1127T} that the SZ 
flux generated by thermal electrons in nearby galaxies could be detected. Note 
that in the previous estimate, we did not include electrons from the very local 
interstellar medium nor from the nearby spiral arm, which we do not expect
to significantly change this approximate result.

\subsection{Dark matter contributions}
\label{subsec:dm_contrib}

\subsubsection{Smooth-halo contribution}
\label{subsubsec:smooth}

In contrast to high-energy electrons of astrophysical origin, 
dark-matter-induced electrons are produced everywhere in the Galactic halo, and 
the relevant line-of-sight length can therefore reach $\sim 100$ kpc. As briefly
mentioned in the Introduction, the very simple exercise of using the local 
astrophysical electron density as a maximum for the dark matter-induced density 
at Galactic radii $r>\rsun=8$ kpc would lead to $\tau_e\lesssim8\times10^{-11}$,
rather far away from current experimental sensitivities. Nevertheless, it is 
worth quantifying more accurately the density distribution of the electrons 
injected by dark matter annihilation (or decay) along the line of sight in the 
high Galactic latitude regions.

The transport of electrons in regions distant from the disk is not very well 
constrained because it is difficult to predict the value of the diffusion 
coefficient. This latter should at least be much larger than locally because 
magnetic turbulences, somehow connected to small-scale inhomogeneities in the 
cosmic-ray plasma, are expected to fade away (\eg~\cite{2009ncrd.book.....S}). 
In any case, the transport of electrons usually relies on a diffusion equation 
which can sometimes be solved in terms of analytical Green functions (see~\eg~
\cite{1964ocr..book.....G,berezinsky_book_90} for extensive reviews, and 
\citesec{subsubsec:diff} for a few further details). In this context, the Green 
function ${\cal G}$ represents the probability for an electron injected at 
position $\vec{x}$ and energy $E_s$ to have propagated to position $\vec{x}$, 
still carrying energy $E\leq E_s$ (we only consider energy losses here), so that
the electron density $dn/dE$ can be expressed in terms of a source ${\cal Q}$ as
follows:
\ben
\frac{dn(E,\vec{x})}{dE}|_{\rm propag} &=& \int_E^\infty dE_s 
\int d^3\vec{x}_s \, {\cal G}(E,\vec{x}\leftarrow E_s,\vec{x}_s)\nn \\
&& \times   {\cal Q}(E_s,\vec{x}_s)\;.
\label{eq:propag_dnde}
\een
As a minimal approach, which will be shown overoptimistic later on, we 
first neglect all processes but the energy losses caused by inverse Compton 
scattering with the CMB photons. If making such a maximal assumption, which 
greatly facilitates the calculation, is not enough to predict an observable SZ 
effect, then rather trustworthy conclusions can easily be drawn. We therefore 
suppose that electrons lose energy at their production site, \ie~we neglect 
spatial diffusion. In that case, referred to as {\em diffusionless} limit
hereafter, the Green function ${\cal G}(E,\vec{x}
\leftarrow E_s,\vec{x}_s) \longrightarrow \delta(\vec{x}_s-\vec{x})/b(E)$, such 
that the electron density at point $\vec{x}$ is related to the local 
annihilation rate as follows:
\ben
\frac{dn}{dE}(E,\vec{x}) \approx
\frac{1}{b(E)} \int_E^{\infty} dE_s \, {\cal Q}_n(E_s,\vec{x})\;,
\label{eq:dnde}
\een
where the source term ${\cal Q}_n$ encodes the dark matter properties as
\ben
{\cal Q}_n(E_s,\vec{x}) = {\cal S}_n 
\left[ \frac{\rho(\vec{x})}{\rho_0} \right]^n\, \frac{dN}{dE_s}\;.
\een
Parameter $\rho_0$ denotes an arbitrary reference density and $dN/dE_s$ is 
the injected electron spectrum. The index $n$ is equal to 2 (1) in the case of 
dark matter annihilation (decay), for which the parameter ${\cal S}$ reads
\begin{align}
{\cal S}_n = 
\begin{cases}
\frac{\Gamma_\chi \, \rho_0}{\mchi} & n=1\\
\delta \frac{\sigv}{2} \left[ \frac{\rho_0}{\mchi} \right]^2 & n=2
\end{cases}\;.
\label{eq:def_s}
\end{align}
We recognize the WIMP mass $\mchi$, the annihilation cross section 
$\sigv$ and the decay rate $\Gamma_\chi$. The parameter $\delta$ is equal to 1 
if annihilation involves identical WIMP particles, or to 1/2 for Dirac fermions.

For simplicity, we first consider annihilation or decay into electron-positron
pairs, so that $dN/dE_s = 2\,\delta(E_s - n\,\mchi/2)$. Indeed, it is clear
from~\citeeq{eq:dnde} that the integrated electron density will mostly depend 
on the total number of electrons produced from dark matter annihilation or 
decay, so it will not be difficult to extrapolate the results obtained with this
specific simple case to other injected spectra (as far as the optical depth is 
the only quantity under investigation and the diffusionless limit is 
considered). From this assumption, we obtain
\ben
n_e(\vec{x}) = {\cal S}_n 
\left[ \frac{\rho(\vec{x})}{\rho_0} \right]^n\, \tau_l \,
{\cal F}(E_{\rm min})\;,
\label{eq:ne_diffless}
\een
where $\tau_l$ is the energy-loss timescale. In the Thomson approximation, the 
energy loss caused by interactions with the CMB is merely given by 
$b(E)=b_0 (\epsilon\equiv E/E_0)^2$ with $b_0= E_0/\tau_l=
2.65\times 10^{-17}\,{\rm GeV/s}$, such that the integrated spectrum ${\cal F}$
can be expressed as
\ben
{\cal F}(E_{\rm min}) =2\,\,E_0\,\left( \frac{1}{E_{\rm min}}
-\frac{2}{n\,\mchi}\right)\simeq \frac{2\,E_0}{E_{\rm min}}\;.
\label{eq:defF}
\een
In the following, we set $E_0$ to 1 GeV, which implies $\tau_l = 3.8\times 
10^{16}$ s. Note that neglecting other sources of energy loss, 
\eg~bremsstrahlung or ionization, is justified far away from the disk where the
interstellar gas density is negligible.

We have now to specify the dark matter mass density shape $\rho(\vec{x})$. While
there are still issues regarding how dark matter concentrates in the centers of 
galaxies, essentially because baryons dominate the central gravitational 
potential in these structures, the off-center regions are less subject to 
debate. Basically, N-body simulations agree on the prediction that the total 
dark matter density (including subhalos) should fall like $\sim r^{-3}$ in the 
outskirts of galaxies, which means that line-of-sight integrals should not 
differ too much among different Galactic halo models towards high-latitude 
regions. In the following, for comparison, we use the results of two recent 
high resolution N-body simulations of Milky-May-like objects, {\em Via Lactea 
II}~\cite{2008Natur.454..735D} and {\em Aquarius}~\cite{2008Natur.456...73S}, in
which the dark matter halos are found well approximated by spherical profiles; a
summary of the relevant ingredients can be found in~\cite{2009arXiv0908.0195P}: 
the former is featured by a $r^{-1}$ Navarro-Frenk-White (NFW) profile
\cite{1997ApJ...490..493N} with a scale radius of $21$ kpc and a local mass 
density of $\rho(\rsun=8\,{\rm kpc}) =  0.42 \,{\rm GeV/cm^3}$, while the latter
follows an Einasto profile with a slope $\alpha = 0.17$, a scale radius of 20 
kpc, and a local density of $0.57\,{\rm GeV/cm^3}$. Though different, these 
local normalizations are in reasonable agreement with the latest constraints to 
date~\cite{2010JCAP...08..004C,2010arXiv1003.3101S}.

In a spherical system centered about the Galactic center with an observer 
located at point $\vec{r}_\odot$, the vector running from the observer along the
line-of-sight $\vec{l}$ and making an angle $\eta'\equiv \pi-\eta$ with 
$\vec{r}_\odot$ can be related to the Galactic radius $r$ through 
$\vec{r}=\vec{r}_\odot + \vec{l}$, such that
\ben
r &=& \sqrt{l^2+r^2_\odot -2\,l\,r_\odot\,\cos\eta}\\
\Leftrightarrow l &=& r_\odot \,\cos\eta + 
\sqrt{r^2-r_\odot^2 \,\sin^2\eta}\;.
\een
This angle $\eta$ can actually be expressed in terms of the pointing angle of 
the telescope $\psi$ with respect to the Galactic center direction, the angle 
$\theta$ that describes the angular resolution, and the angle $\phi$ that 
runs circularly around the pointing direction,
\ben
\cos\eta = \sin\psi\,\sin\theta\,\cos\phi+\cos\theta\,\cos\psi\;.
\een
Note that because of the spherical symmetry, $\psi$ merely corresponds to 
(minus) the Galactic latitude $b$ for an observer located on Earth, as measured 
at longitude $0^\circ$. Armed with these relations, we can readily compute 
the electron density at a given position $l$ along the line of sight, such that 
the optical depth averaged over the resolution angle $\theta_{\rm res}$ defined
in \citeeq{eq:def_tau_av} is finally given by
\ben
\langle \tau_{e,n}^{\chi}(\psi) \rangle_{\rm res} = \sigma_{\rm T} 
{\cal S}_n \, \tau_l \, {\cal F}(E_{\rm min}) \,
 \left( 2\,r_0\,{\cal J}_n(\psi,\mu_{\rm res})\right)\;,
\label{eq:tau_chi_ave}
\een
where $\mu_{\rm res} = \cos(\theta_{\rm res})$. Similarly to what is encountered 
in indirect dark matter detection with gamma rays~\cite{1998APh.....9..137B}, 
we have defined the dimensionless parameter ${\cal J}$, the averaged 
line-of-sight integral, as follows:
\begin{align}
\label{eq:def_j}
{\cal J}_n(\psi,\mu_{\rm res}) &\equiv
\frac{1}{4\,\pi\,r_0(1-\mu_{\rm res})} \, \times \\
& \int_0^{2\pi}d\phi \int_{\mu_{\rm res}}^{1}d\mu \, \int_0^{\infty}dl\,
\left[\frac{\rho(l(r,\psi,\mu,\phi))}{\rho_0}\right]^n \nn \;.
\end{align}
We will further use $\rho_0=0.3\,{\rm GeV/cm^3}$ and $r_0=r_\odot=8$ kpc.

\begin{figure}[t!]%[htp]
 \centering
\includegraphics[width=\columnwidth]{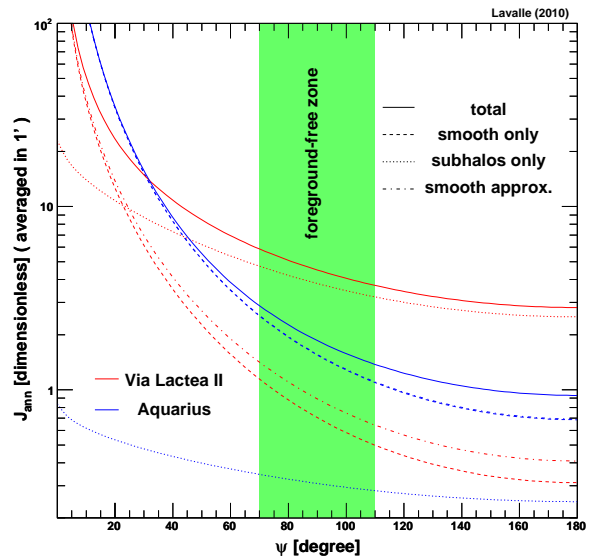}
\caption{Plot of the parameter ${\cal J}_{\rm ann}\equiv J_2$ corresponding to 
  the dark matter annihilation case [see \citeeq{eq:def_j}] for two 
  configurations, the first one inspired from {\em Via Lactea II}
  \cite{2008Natur.454..735D} and the other one inspired from 
  {\em Aquarius}~\cite{2008MNRAS.391.1685S}.
  The dashed curves correspond to the smooth only contributions, the dash-dotted
  curves to the smooth approximation, the dotted curves to the subhalo 
  contributions, and the solid curves the sum of the smooth-only and subhalo 
  contributions.}
\label{fig:jpsi}
\end{figure}

\begin{figure}[t!]%[htp]
 \centering
\includegraphics[width=\columnwidth]{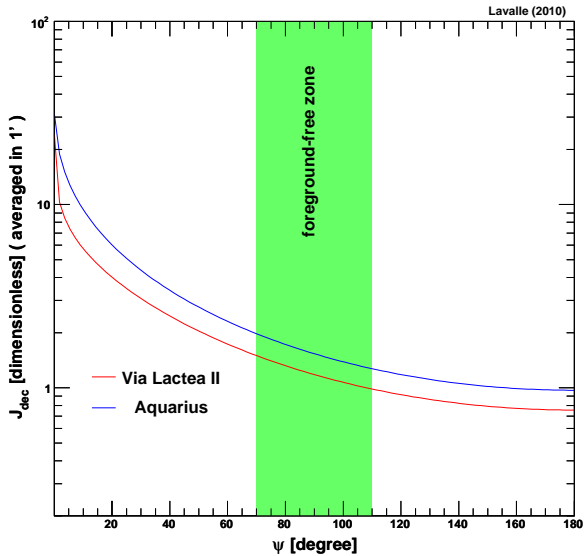}
\caption{Plot of the parameter ${\cal J}_{\rm dec}\equiv J_1$ corresponding to 
  the dark matter decay case (see \citeeqp{eq:def_j}) for the same two configurations
  as in \citefig{fig:jpsi}. For dark matter decay, subhalos do not play any role in
  average.}
\label{fig:jpsi_dec}
\end{figure}

The results obtained for the numerical calculation of ${\cal J}_2 (\psi,
\theta_{\rm res}=1')$ are displayed in \citefig{fig:jpsi}. The high-latitude 
zone corresponds to $70^\circ \lesssim \psi \lesssim 110^\circ$ (the colored 
region in the plot), for which the astrophysical foreground is expected to be 
the lowest. The dash-dotted curves represent the so-called {\em smooth 
approximation} for both the dark matter profiles discussed above, \ie\ 
disregarding the potential presence of subhalos --- their impact is discussed 
in \citesec{subsubsec:sub}. The parameter ${\cal J}_2$ associated with the 
smooth dark matter halo is found $\lesssim 2$ in the region of interest 
(shaded). We find similar results on large latitudes for the dark matter decay 
case, as illustrated in \citefig{fig:jpsi_dec}.

To summarize, we provide a crude estimate of the optical depth in the smooth 
approximation. If we define
\ben
\label{eq:nchi}
\bar{\cal N}^\chi_n &\equiv&
\frac{\tau_l}{3.8\times 10^{16}{\rm s}} 
\, \frac{{\cal S}_n }{1.35\times 10^{-27}{\rm cm^{-3}s^{-1}}} \\
&& \times \frac{{\cal F}(m_e) }{2/(511\times 10^{-6})}
\, \frac{r_0\times {\cal J}_n }{8\,{\rm kpc}\times 2} \;,\nn
\een
which is equal to $ \delta $ --- \ie\ 1 or 1/2 [see \citeeq{eq:def_s}] 
--- (or 1) for the canonical values of the parameters made explicit 
above in the case of dark matter annihilation (decay, respectively). The 
reference value given above for ${\cal S}_n$ was obtained by assuming 
$\sigv = 3\times 10^{-26}{\rm cm^3/s}$ (n=2, annihilation) or $\Gamma_\chi = 
4.5\times 10^{-27}{\rm s^{-1}}$ (n=1, decay)\footnotemark, for $\mchi = 1$ GeV 
and $\rho_0 = 0.3\,{\rm GeV/cm^3}$ [see \citeeq{eq:def_s}]. The average
optical depth is then approximately
\ben
\langle \tau_{e,n}^{\chi}(90^\circ) \rangle_{1'} = 
1.3\,10^{-8}\bar{\cal N}^\chi_n \;.
\label{eq:av_taudm}
\een
This value, which roughly sizes the amplitude of the SZ signal expected from 
high latitudes, is very small, much smaller than the typical optical 
depth found for Galactic thermal electrons and, more importantly, than the 
current experimental sensitivities. Indeed, we recall that new generation 
experiments, like Planck~\cite{2006astro.ph..4069T} or ALMA
\cite{alma_paper,2009IEEEP..97.1463W}, can hardly constrain electron populations
of optical depths $\lesssim 10^{-4}$. We emphasize that we used a rather light 
WIMP mass of 1 GeV in this estimate 
(see \eg~\cite{2004NuPhB.683..219B,2004PhRvD..69j1302B} for more detailed 
phenomenological aspects on light dark matter) and considered the favorable case
of a diffusionless ``transport'' for electrons. Likewise, we employed a value 
of ${\cal F}$ down to the rest energy of electrons, and a reasonable angular 
resolution of $1'$ --- the angular resolution does not play a significant role 
when the line of sight is offset from the center of the target structure. These 
optimal assumptions still lead to a weak result, comparable with what we 
obtained for the relativistic astrophysical foreground but much smaller than
the thermal one [see \citeeqss{eq:tau_bg}{eq:tau_bg_th}], which makes the SZ 
effect a too feeble tracer of dark matter annihilation or decay for a smooth
Galactic halo.

\footnotetext{The decay rate cannot be much larger than this
  value to obey the diffuse gamma-ray constraint 
$\Gamma_\chi\lesssim 10^{-26}$ obtained in~\cite{2010JCAP...06..027Z}.}

A larger amplitude could be reached with lighter dark matter particles, but 
additional astrophysical constraints, \eg\ coming from hard x-ray observations 
of the Galactic center~\cite{2005A&A...441..513K}, may then bound the 
annihilation cross section to smaller values as well
\cite{2004PhRvL..92j1301B,2006MNRAS.368.1695A}. The presence of dark matter 
substructures could also increase the amplitude in the case of dark matter 
annihilation, which is precisely the topic of the next paragraph. Note that for 
dark matter decay, subhalos, which are small-scale inhomogeneities, are not 
expected to significantly boost the SZ signal because the decay rate scales 
linearly with the dark matter density: the above smooth-halo approximation is 
likely a rather good approximation in that case.

\subsubsection{Subhalo contributions}
\label{subsubsec:sub}
So far, we have considered a spherical and smooth dark matter halo without 
discussing the role of subhalos (see \citesec{subsubsec:sub}). In this
section, we study two different cases: (i) the collective effect of a subhalo 
population and (ii) the impact of a single big subhalo located along the line 
of sight. We recall that subhalos are expected to play a more minor role in the 
case of dark matter decay.

\subsubsection{Average subhalo contribution}

Let us first assume that subhalos contribute another smooth injection
of electrons, the rate of which is set by their inner properties averaged over
their spatial and mass distributions. Subhalos are indeed usually described in 
terms of (i) their global properties, \ie\ their mass and spatial distributions,
and (ii) their inner properties, \ie\ their mass $m$, mass density shape 
$\rho_{\rm sh}$, concentration $c$ and position $r$ in the host halo. 
Theoretical prescriptions can be found for both types from cosmological 
simulation results. Once the subhalo properties are fixed, the global 
associated annihilation rate can be calculated. Thus, the dimensionless 
${\cal J}$ factor associated with the electron injection from a population of 
subhalos is given by
\begin{align}
{\cal J}_{\rm sh}(\psi) & = 
\frac{N_{\rm sh}^{\rm tot}}{4\,\pi\,r_0(1-\mu_{\rm res})} \, \times \\
\int_0^{2\pi} & d\phi \int_{\mu_{\rm res}}^{1}d\mu \, \int_0^{\infty}dl\,
\langle \xi (l)\rangle_m \frac{d{\cal P}(l(r,\psi,\mu,\phi))}{dV}\;,\nn
\end{align}
where $N_{\rm sh}^{\rm tot}$ is the total number of subhalos in the Milky Way, 
$d{\cal P}/dV$ is the spatial probability distribution function (pdf), and
\ben
\langle \xi (l)\rangle_m \equiv \int dm \frac{d{\cal P}}{dm}
\int d^3\vec{x}_{\rm sh}
\left[ \frac{\rho_{\rm sh}(\vec{x}_{\rm sh},r,m,c)}{\rho_0} \right]^2
\een
is proportional to the mean subhalo annihilation rate ($d{\cal P}/dm$ is the 
mass pdf).

Using this smooth approximation for a subhalo population implies making the 
assumption that the electron density carried inside the angular resolution of 
the telescope does not fluctuate. It is therefore more appropriate
for large-index mass pdfs (scaling typically between $M^{-1.9}$ and $M^{-2}$) 
that favor the relative contribution of the lightest subhalos, which are also 
the smallest, the most concentrated and the most numerous --- see 
\cite{2008A&A...479..427L} for more details on the influence of the subhalo 
parameters.

The impact of considering the average subhalo contribution is reported 
in~\citefig{fig:jpsi}, where we used the {\em Via Lactea II} and {\em Aquarius}
subhalo phase spaces defined in \cite{2009arXiv0908.0195P}, for which the mass 
functions scale like $M^{-2}$ and $M^{-1.9}$, respectively --- a free-streaming 
cutoff of $10^{-6}\msun$ is taken. The dashed curves correspond to the 
contribution of the smooth host halos only --- different from the {\em smooth 
approximation} studied in \citesec{subsubsec:smooth} (dash-dotted curves) 
because part of the dark matter mass is now in the form of subhalos. We note 
that the smooth approximation and the smooth-only contribution are almost 
superimposed in the case of the {\em Aquarius} model, which is partly due to the
fact that subhalo mass fraction is much smaller in this setup (17\%) than in the
{\em Via Lactea II} one (51\%); the slightly disadvantageous mass distribution 
and internal subhalo structure also contribute to diminish the impact of 
subhalos 
in the {\em Aquarius} case. The values obtained for ${\cal J}_{\rm sh}(\psi)$,
namely the average subhalo contributions, appear as dotted curves. We see that 
in the shaded foreground-free zone, the subhalo contribution of the {\em Via 
Lactea II}
model exceeds the smooth host halo one by half an order of magnitude, reaching 
${\cal J}_{\rm sh}(90^\circ)\approx 5$. In contrast, the {\em Aquarius} subhalo 
configuration leads to an average signal lower than the smooth host halo one, 
lying 1 order of magnitude below what is obtained for {\em Via Lactea II}. 
Such a difference mostly comes from the larger mass index of 2 taken in the 
{\em Via Lactea II} setup, which results in more mass in the form of subhalos, 
and favors the relative contribution of smaller and more concentrated objects. 
This gap between these two dark matter modelings provides an idea of the 
average theoretical uncertainties affecting the predictions involving subhalos.

In summary, it seems that subhalos, when taken globally, can lead to an average 
SZ signal enhancement up to a factor of $\sim 5$ in the high-latitude 
predictions, which, by means of \citeeq{eq:av_taudm} corresponds to an optical 
depth of $\langle\tau^\chi_n\rangle\sim 5\times 10^{-8}$, still far below 
experimental sensitivities. We also emphasize that the theoretical prescriptions
that we employed here for substructures are inferred from dark-matter-only 
cosmological simulations. We could expect the Galactic baryonic disk and bulge 
to further decrease the subhalo impact due to more efficient tidal stripping. 
We address the case of single objects in the next paragraph.

\subsubsection{Individual subhalo contributions}
\label{subsubsec:single_subh}

\begin{figure}[t!]%[htp]
 \centering
\includegraphics[width=\columnwidth]{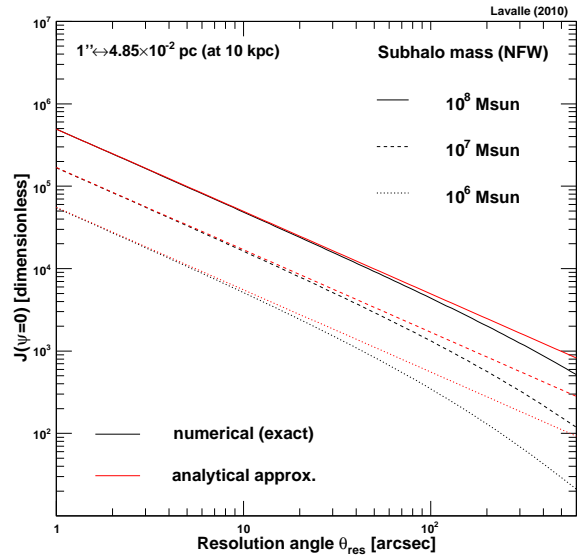}
\caption{The dimensionless ${\cal J}_{\rm single}$ factors for three different 
  subhalo masses, taken in the range $10^6-10^8\msun$, as functions of the 
  resolution angle $\theta_{\rm res}$. The analytical approximation proposed
  in \cite{2010JCAP...02..005L} is also shown. All objects are assumed to be 
  located at 10 kpc from the observer. This factor is computed using 
  $\psi = 0^\circ$, where $\psi$ refers to the angle of the line of sight with 
  the subhalo center.}
\label{fig:subhalo}
\end{figure}

To study the contribution of isolated objects, we consider three subhalos within
the mass range $10^6$-$10^8$ \msun\ and with 
inner NFW density profiles --- adopting instead an Einasto shape would not
change the final results. Their radial extents $r_{200}$ are connected to 
their scale radii $r_s$ via the concentration parameter $c_{200}$ through the 
relation $c_{200} = r_{200}/ r_s$. The parameters that we use are reported in
\citetab{tab:cl}, where the angular sizes of the scale radii are given assuming
a distance to the observer of 10 kpc for all objects. These parameters are 
close to those inferred from the {\em Via Lactea II} setup used in 
\cite{2009arXiv0908.0195P}, to which we refer the reader for 
more details, and roughly correspond to what can be expected for subhalos 
located at $\sim 10$ kpc from the Galactic center (the closer to the Galactic
center, the more concentrated).

\begin{table}
\centering
\begin{tabular}{ccccc}
\hline
Subhalo mass & $r_{200}$ & $c_{200}$  & $r_s$& $\rho_s$ \\
$[\msun]$ & $[{\rm kpc}]$ & & $[{\rm pc}\, (')^\dagger]$ &  $[{\rm GeV/cm^3}]$ \\
\hline
$10^6$ & $2.01$ & $56.7$ & $35.4$ $(12.2')$  & $22.2$ \\
$10^7$ & $4.32$ & $50.0$ & $86.4$ $(29.7')$  & $15.9$ \\
$10^8$ & $9.31$ & $43.5$ & $213.9$ $(73.5')$ & $11.0$ \\
\hline
\end{tabular}
\caption{NFW parameters of the subhalos used for the isolated source analysis.
  $^\dagger$ The angular size associated with the scale radius $r_s$ is derived 
  assuming that the object is located at a distance of 10 kpc from the 
  observer.}
\label{tab:cl}
\end{table}

From \citetab{tab:cl}, we can already notice that, since most of the 
annihilation should occur within the scale radius of an NFW target, all objects 
look extended to any (sub)arcmin-resolution experiment when assuming a distance 
of 10 kpc and a diffusionless transport for electrons. Since the scale 
radius roughly scales like $M^{1/3}$, the biggest resolved objects have masses 
$\lesssim 10^{3}\msun$ at this distance. Interestingly enough, the lower mass 
range down to $10^{-6}\msun$ is the one that dominates the overall subhalo 
contribution for a pdf mass index larger than 1.9 \cite{2008A&A...479..427L} 
--- it is equal to 2 in the {\em Via Lactea II} model adopted here --- which 
itself bypasses the smooth host halo yield. Therefore, treating apart big 
subhalos as we do here has no consequence on the average subhalo contribution
that we worked out earlier, which does not have to be depleted. As we will see
later, however, the subhalo size does not reflect, in fact, the actual electron 
distribution extent associated with the object, due to spatial diffusion 
effects.

It is clear that though extended, our template single subhalos should increase
the detection potential if crossing the line of sight of the telescope, 
because of the dark matter cusps they contain. To estimate the corresponding 
optical depths, we can first calculate the ${\cal J}_{\rm single} = 
{\cal J}_2(\psi = 0^\circ)$ factor defined in \citeeq{eq:def_j} for each object,
which implicitly corresponds to using the diffusionless limit. We show our 
results in \citefig{fig:subhalo}, where ${\cal J}_{\rm single}$ is plotted 
against the angular resolution. We see that for an angular resolution around 
1', values of $\sim 10^3$-$10^4$ can be reached for a big subhalo located at a 
distance of 10 kpc. By using \citeeqss{eq:nchi}{eq:av_taudm}~and parameters 
therein, this translates into an optical depth of $\langle\tau^\chi_n\rangle
\sim 10^{-5}$-$10^{-4}$. Note that since the angular extent of the scale radii 
of our objects is larger than the reference angular resolution of 1', we can 
use the {\em line-of-sight approximation} of ${\cal J}_2$ proposed in 
Ref.~\cite{2010JCAP...02..005L}, which is analytical (denoted 
${\cal J}_{\rm los}^2$ in their Eq.~3.20). The relation between both formalisms 
is the following:
\ben
\label{eq:j_approx}
{\cal J}_2(\psi=0^\circ) &\approx& 
\frac{r_s}{r_0}\,\left[\frac{\rho_s}{\rho_0}\right]^2\, {\cal J}_{\rm los}^2 \\
&\approx& \left[\frac{\rho_s}{\rho_0}\right]^2\, 
\frac{(1+\mu_{\rm res})\,\pi}{2}\frac{r_s^2}{r_0\,b_{\rm res}}\nn\;,
\een
where $b_{\rm res}\equiv d\,\sin\theta_{\rm res}$ is the impact parameter.
This approximation is demonstrated to be very accurate in 
\citefig{fig:subhalo} except when the angular resolution exceeds the 
subhalo scale radius. We recall that the main assumption behind this 
formula is to approximate an inner profile to its central behavior namely 
$\rho(r)\approx \rho_s (r/r_s)^{-\gamma}$.

Such a result, \ie~$\langle\tau^\chi_n\rangle\sim 10^{-5}$-$10^{-4}$, might fall 
within the current experimental sensitivities and is therefore worth being more 
deeply investigated --- note, however, that it is obtained for a quite light 
WIMP of 1 GeV. In particular, it is important to study the additional and 
fundamental role of spatial diffusion.

\begin{figure*}[t!]%[htp]
 \centering
\includegraphics[width=0.68\columnwidth]{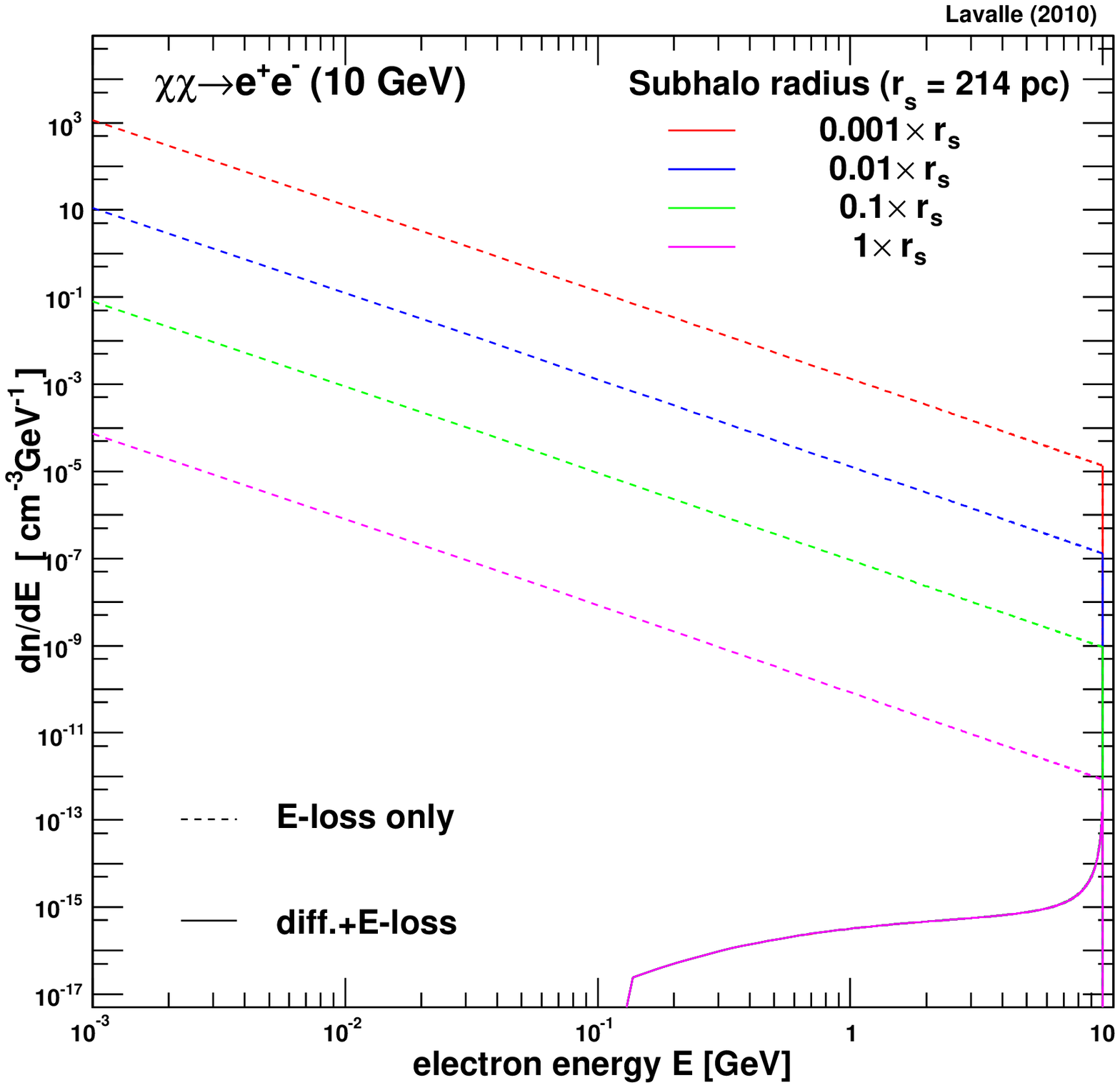}
\includegraphics[width=0.68\columnwidth]{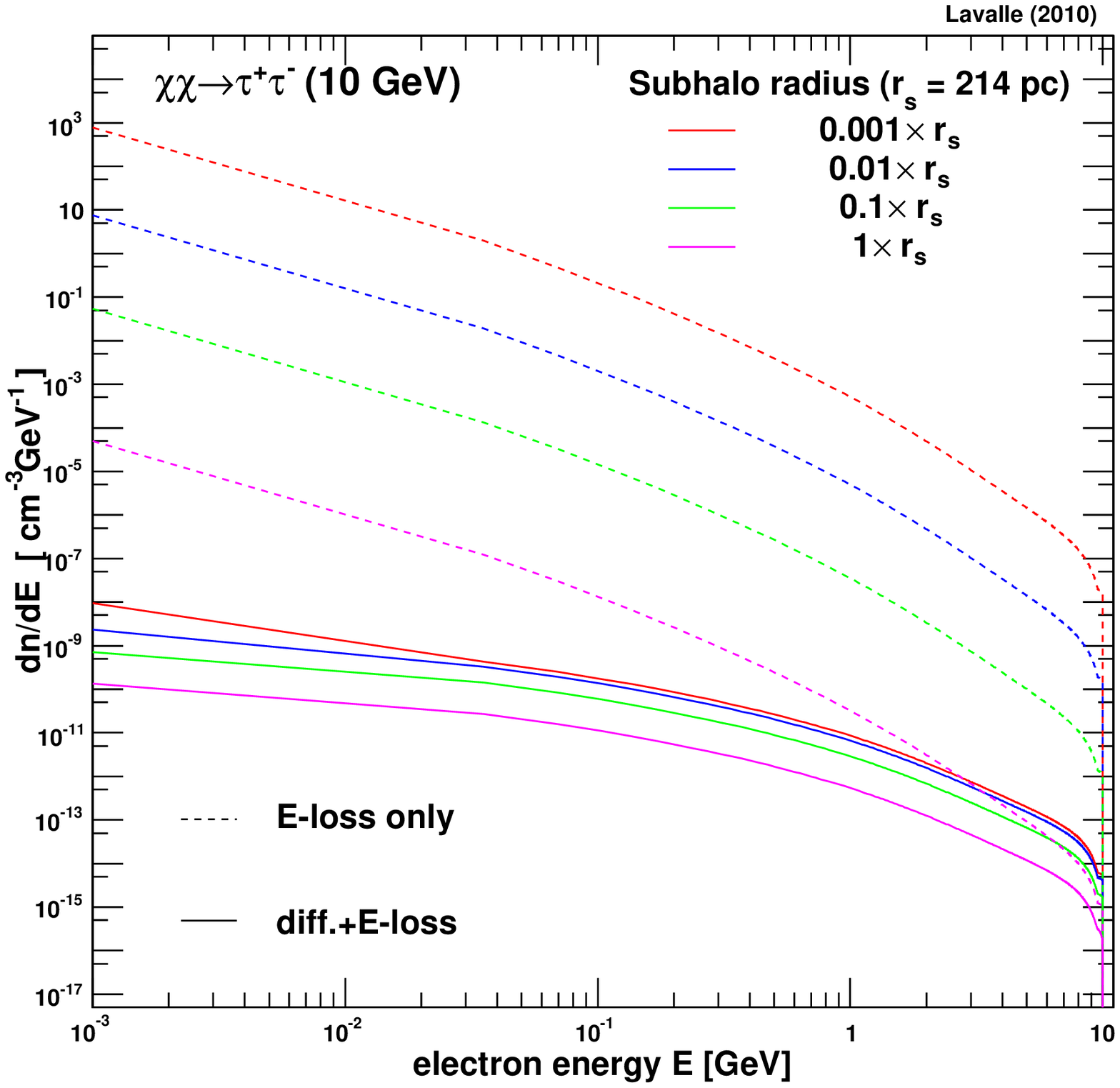}
\includegraphics[width=0.68\columnwidth]{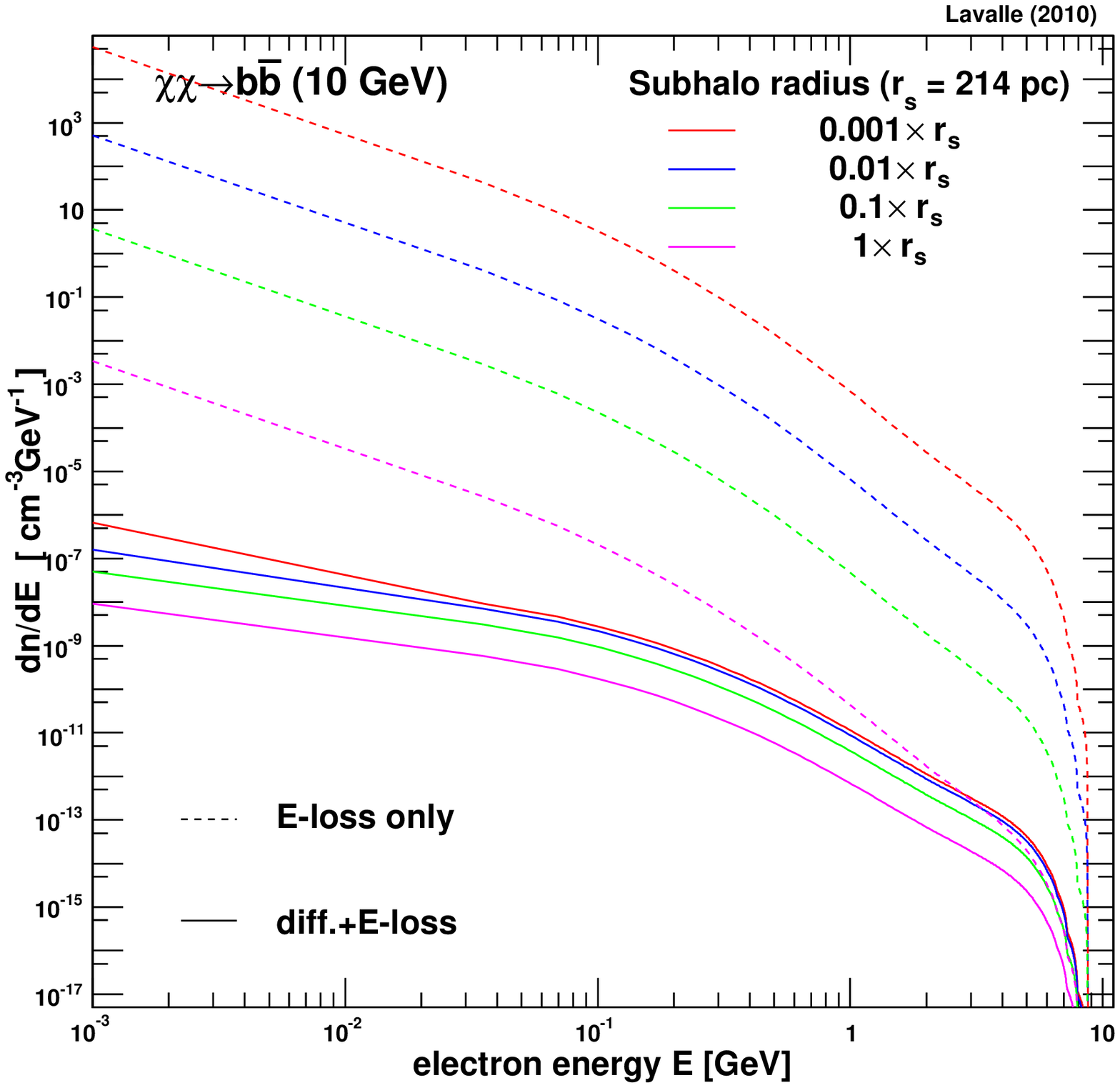}
\caption{Left: comparison of the electron density obtained in a full diffusion
  model, at different positions in the $10^8\msun$ subhalo of~\citetab{tab:cl}, 
  with the electron density calculated in the diffusionless approximation; a 
  monochromatic injection of 10 GeV electrons is assumed. Middle (Right): same 
  as left panel, but assuming a $\tau^+\tau^-$ ($b\bar{b}$) injection spectrum.}
\label{fig:diff_ne_sub}
\end{figure*}

\subsubsection{A focus on spatial diffusion effects}
\label{subsubsec:diff}

Our most critical assumption so far was to neglect the spatial diffusion of 
electrons, so it is first interesting to compare the relevant spatial scales. 
An angular size of 1' corresponds to a physical of $\sim 3$ pc for a target 
located at 10 kpc, scaling almost linearly with the distance. If we write the 
diffusion coefficient as $K(E)\equiv K_0 (E/E_0)^\delta$ (see 
\eg~\cite{berezinsky_book_90}), then the electron propagation scale $\lambda$ 
can be defined in a steady-state regime as
\ben
\label{eq:approx_lambda}
\lambda^2(E\leftarrow E_s) &\equiv& 4\int_E^{E_s} dE'\frac{K(E')}{b(E')}\\
&=& 4 \,K_0 \,\tau_l \frac{(E_0/E)^{1-\delta}}{(1-\delta)}\left[1- 
\left(\frac{E}{E_s}\right)^{1-\delta}\right]\nn\;,
\een
where $b(E)$ is the energy-loss rate (taken in the Thomson approximation), 
$E_s$ the injected electron energy and $E<E_s$ the energy after the electron has
propagated over a distance of $\lambda>0$ on average. It is difficult to 
determine the diffusion coefficient far away from the Galactic disk because most
of observational constraints are local (see \eg~\cite{2001ApJ...555..585M} 
or~\cite{1998ApJ...509..212S}). Nevertheless, we can consider the local value 
as a lower bound, since diffusion is expected to be more efficient in a less 
dense and less turbulent medium~\cite{2009ncrd.book.....S} --- the densities of 
interstellar matter and cosmic rays are expected (and observed) to decrease 
with the distance to the disk. For simplicity, let us assume that $K_0=3\times 
10^{27}\,{\rm cm^2/s}$ ($E_0=1$ GeV) and $\delta=0.7$, values close to those 
inferred from local constraints 
(see~\eg~\cite{2001ApJ...555..585M,2010A&A...516A..66P}).
Further supposing, as before, that the energy-loss rate is only driven by 
interactions with the CMB, we have,
\ben
\lambda(E\leftarrow E_s) & = & 12.6\,{\rm kpc}\,\\
&& \times
\left\{ \left[\frac{1\,{\rm GeV}}{E}\right]^{0.3}- 
\left[\frac{1\,{\rm GeV}}{E_S}\right]^{0.3}\right\}^{1/2}\;.\nn
\een
Thus, if we consider an injected energy of $E_s=1$ GeV, the propagation scale 
becomes larger than $\sim 3$ pc for $\delta E \equiv E_s-E > 2\times 10^{-7}$ 
GeV. This tremendously small value of $\delta E$ actually defines the spectral 
domain of validity of our previous estimate of the optical depth, when spatial 
diffusion was neglected. The actual electron density traces the squared dark 
matter density in subhalos at the very moment of injection, and is smeared out
afterwards due to propagation effects, which induces the formation of a core of 
electrons. By comparing the scales, it is clear that
diffusion effects completely overcome angular resolution effects: the
propagation scale derived above is as large as, or even larger than, a big 
subhalo itself. The fact that the smearing due to propagation dominates angular 
resolution effects was already emphasized in the context of galaxy clusters in 
Ref.~\cite{2010JCAP...02..005L}, and in the context of dwarf spheroidal 
galaxies in Refs.~\cite{2007PhRvD..75b3513C,2010arXiv1005.2325H}. More 
generally, smearing effects are important whenever the source gradient is large
over a typical diffusion length.

To more precisely illustrate the role of spatial diffusion, we adopt a very 
simple three-dimensional isotropic and homogeneous diffusion model which is 
defined by the steady-state equation
\ben
-K(E)\,\Delta {\cal N}(E,\vec{x}) - \partial_E \left\{ b(E)\,{\cal N}(E,\vec{x})
\right\} = {\cal Q}(E,\vec{x})\;,\nn\\
\een
for which the Green function is analytical:
\ben
{\cal G}(E,\vec{x}\leftarrow E_s,\vec{x}_s) = 
\frac{\exp\left\{-\frac{\left( \vec{x} - \vec{x}_s\right)^2}{\lambda^2}\right\}}
{b(E)\left[\pi\,\lambda^2\right]^{3/2}}
 \;.
\een
Supposing a still subhalo, the propagated equilibrium electron density at 
position $\vec{x}$ in the subhalo is therefore given by plugging the previous 
Green function into \citeeq{eq:propag_dnde}.

In~\citefig{fig:diff_ne_sub}, we report the equilibrium electron density 
calculated for different positions in the $10^8\msun$ subhalo featured 
in~\citetab{tab:cl}, obtained both (i) in the diffusionless approximation 
(dashed curves) and (ii) in a full diffusion-loss propagation model (solid 
curves). A direct annihilation into electron-positron pairs is assumed in the 
left panel ($E_s = \mchi$), and for completeness, we also consider the case of 
a $\tau^+\tau^-$ ($b\bar{b}$) annihilation spectrum in the middle (right) panel.
For each case, we take a WIMP mass of 10 GeV and the canonical value for the 
annihilation cross section, and we compute the electron density at different 
positions in
the subhalo in between a thousandth of $r_s$ and $r_s$, \ie~in the bulk of the 
injection region. We stress that a 10 GeV WIMP annihilating into $b\bar{b}$ 
pairs is already likely excluded by cosmic-ray antiproton data
\cite{2010arXiv1007.5253L}, but such an example is still useful in terms of 
spectral properties. In the left panel, the diffusionless approximation is 
demonstrated to tend towards the full calculation only in the limit $\delta E
\rightarrow 0$, as expected, while the discrepancy is shown dramatic over the 
remaining --- \ie~almost entire --- part of the spectrum. We notably see that 
below $\sim 200$ MeV, all electrons have diffused away from the subhalo, and 
that their density is almost constant all over it for higher energies --- 
except for $\delta E\rightarrow 0$. This can be understood 
from~\citeeq{eq:approx_lambda}: for $E_s=10$ GeV, $\lambda$ becomes larger than 
$r_s = 214$ pc when $\delta E \gtrsim 0.1$ GeV, which results into smearing the 
differential electron density over that scale, setting a cored spatial 
distribution over most of the spectrum. In this pair-injection case, the 
diffusionless approximation can lead to an overestimate of the electron density
by more than 3 orders of magnitude, a discrepancy that strongly increases from 
the edge of $r_s$ to the very central region of the subhalo. The error is 
slightly decreased when considering a continuous $\tau^+\tau^-$ or $b\bar{b}$ 
spectrum because injection proceeds at any energy less than the WIMP mass. In 
that case, it still amounts to a few orders of magnitude, increasing when energy
decreases. We also note that though the annihilation rate varies by a factor
of $\sim 4,000^2$ (NFW case) between $r_s$ and $0.001\,r_s$, the differential 
electron density only spans a bit less than 2 orders of magnitude in the 
$\tau^+\tau^-$ or $b\bar{b}$ case, which shows that diffusion is a very 
efficient spatial smearing process. It is therefore clear that accounting for 
spatial diffusion will rescale the optical depth to much lower values than
estimated before in the case of individual massive subhalos.

In \citefig{fig:diff_ne_sub}, we have assumed $K_0=3\times 
10^{27}\,{\rm cm^2/s}$
for spatial diffusion. Such a value, which is constrained locally, is not 
expected to be relevant to regions distant by more than a few kpc from the 
Galactic plane, where diffusion should become closer and
closer to free propagation. Nevertheless, we emphasize that using a more 
realistic value will not qualitatively change our argument about spatial 
diffusion. Indeed, a more realistic value for the $K_0$, though 
still to be determined, should at least be larger than the local one, 
and would therefore lead to a larger propagation scale for electrons, which
even strengthens the diffusion effect.

%%% Re-reading stopped here. JL

\begin{figure*}[t!]%[htp]
 \centering
\includegraphics[width=\columnwidth]{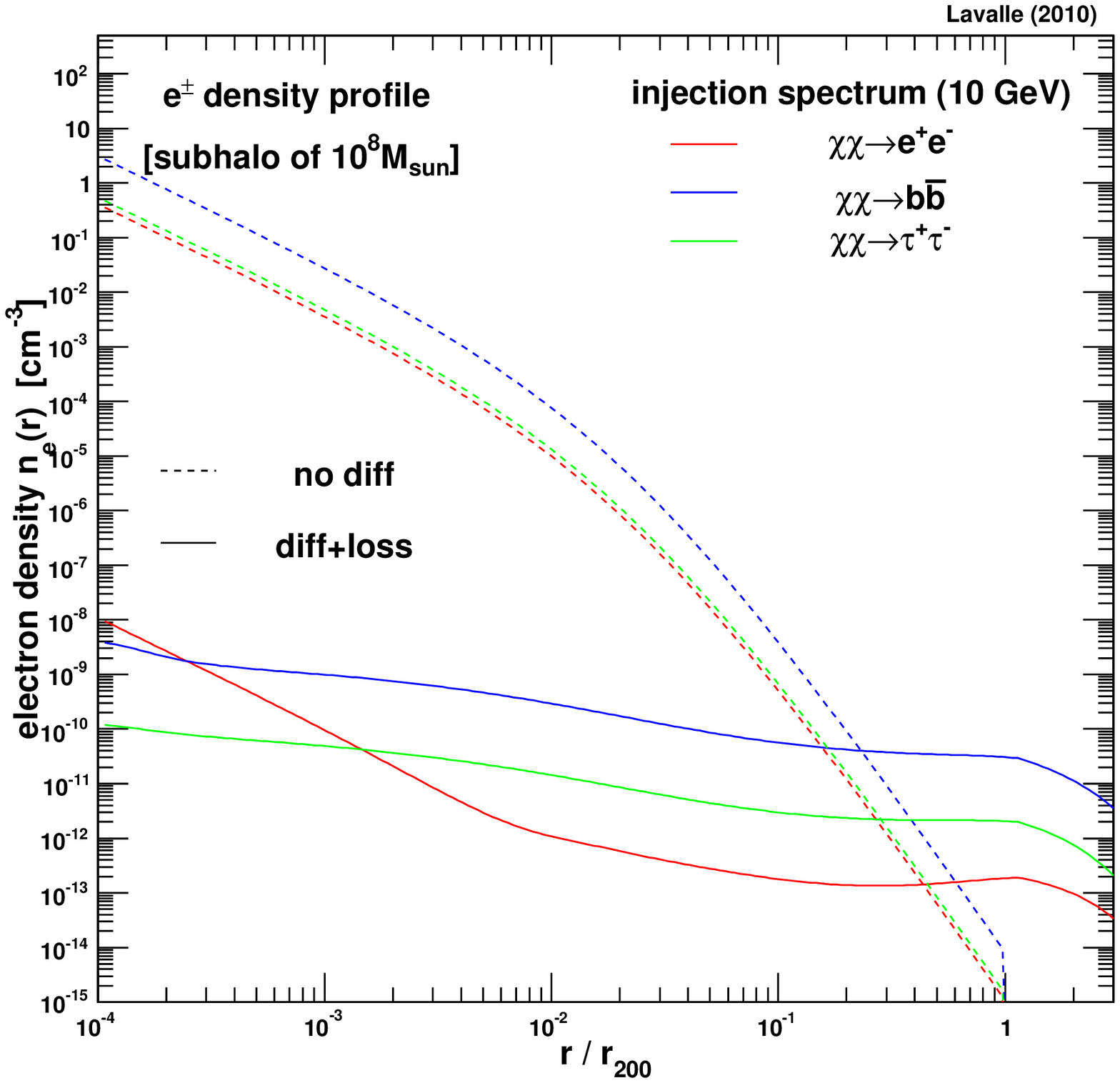}
\includegraphics[width=\columnwidth]{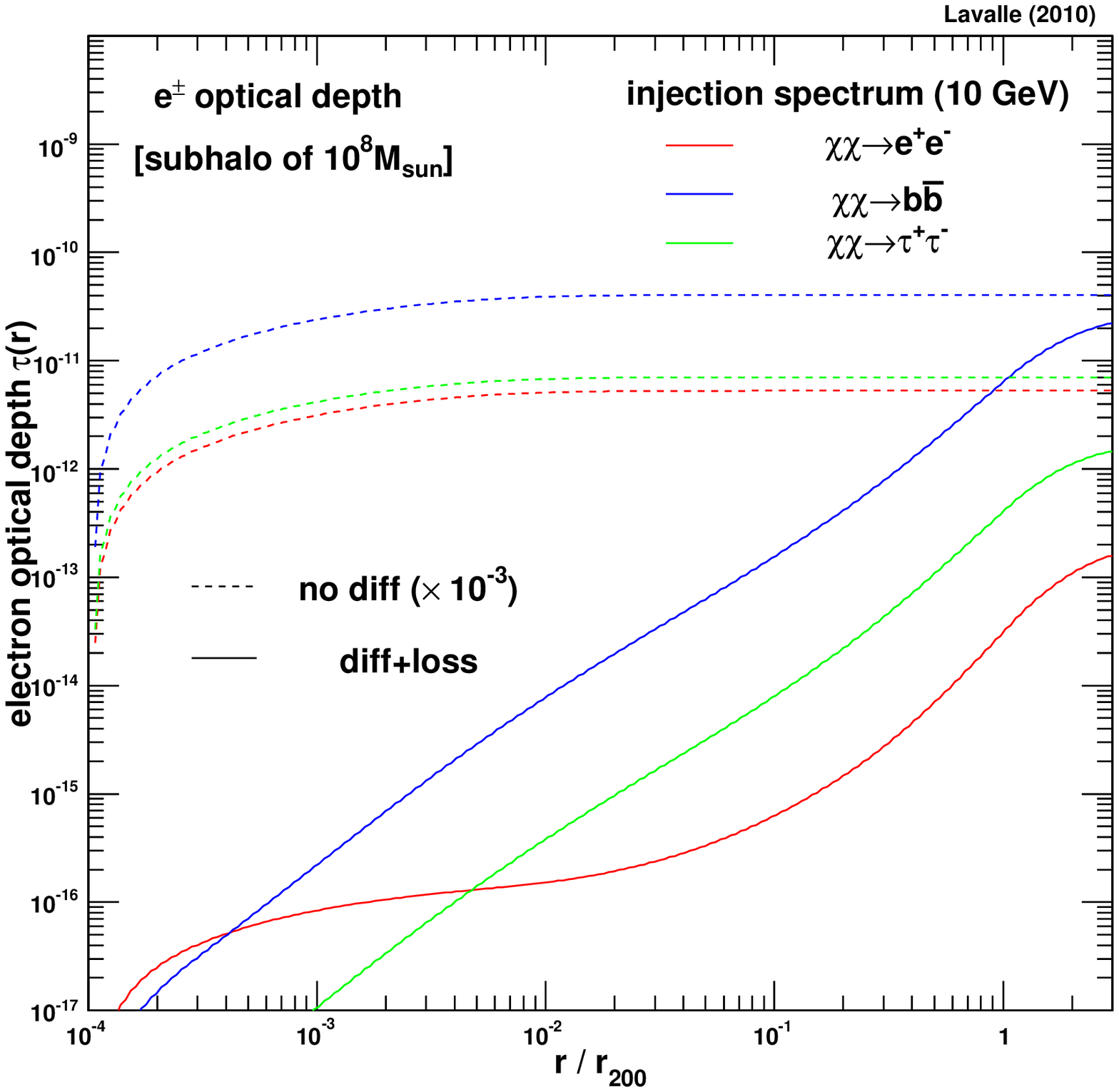}
\caption{Left: Electron density as a function of the subhalo radius for 
  different injection spectra. Right: Corresponding optical depth against
  the subhalo radius ($r=l/2$, where $l$ is the line-of-sight length). In
  both panels an annihilating 10 GeV WIMP is assumed; dashed curves represent
  the diffusionless case, while solid curves refer to a full diffusion-loss
  transport model.}
\label{fig:ne_tot_sub}
\end{figure*}

We can now go further in the calculation of the optical depth by adopting
the full diffusion-loss transport model described above. In 
\citefig{fig:ne_tot_sub}, we derive the electron density profile associated
with our $10^8$ subhalo (left panel) and the corresponding cumulative optical 
depth (right panel). We again consider a 10 GeV WIMP annihilating into the three
different final states discussed above and illustrate the discrepancy
between a full diffusion-loss transport model (solid curves) and the 
diffusionless approximation (dashed curves). In the left panel, the electron
density as derived in the full transport model exhibits a quasicore up to the 
subhalo extent $r_{200}=9.3$ kpc, which is incidentally of the same order of 
$\lambda$, whereas the diffusionless density scales completely differently like
$\rho^2(r)\propto r^{-2}$ --- in the full transport model, the electron profile
extends beyond the subhalo itself. This has important consequences for the 
optical depth (right panel), which is the line-of-sight integral of the density,
since it increases linearly with the radius up to $r_{200}$ in the former case, 
while logarithmically in the latter case. The total optical depth can be read 
off at the border of the object, and we see that the discrepancy between the two
transport hypotheses lies within 3 to 5 orders of magnitude, from soft to hard 
spectral properties (the diffusionless curves are rescaled by a factor of 
$10^{-3}$). Finally, we see that for such a massive subhalo and for a 10 GeV
WIMP with canonical properties, the optical depth $\tau_e < 10^{-11}$, far away
from experimental sensitivities. Going to lower WIMP masses would favor the
annihilation into lepton-antilepton pairs and would therefore not benefit 
of the favorable $1/ \mchi^2$ factor as optimally as necessary.

To summarize this section, we find that, considering the GeV mass scale 
for WIMPs, subhalos are not expected to provide an observable SZ contribution 
due to the very weak optical depth they generate --- of the order of $\tau_e 
\lesssim 10^{-9}$ collectively down to $\tau_e \lesssim 10^{-11}$ individually, 
for a 10 GeV annihilating particle. In the former case, our calculations were 
derived in the optimistic diffusionless limit but still led to pessimistic 
values. In the latter case, spatial diffusion was shown to have the most 
dramatic impact on predictions because it dilutes away the electron density 
injected at high rate at subhalo centers. This is in agreement with the 
pessimistic results found in the context of dwarf spheroidal galaxies 
\cite{2007PhRvD..75b3513C,2010arXiv1005.2325H}, which can be considered as
very massive subhalos, though with a sizable baryon fraction inside --- the
energy-loss rate of electrons is then driven by ionization at low energy.
For annihilating dark matter, decreasing the WIMP mass down to the MeV scale 
would increase the electron injection rate by 6 orders of magnitude if one 
merely considers the favorable $1/\mchi^2$ scaling relation. Nevertheless, we 
see from the right panel of~\citefig{fig:ne_tot_sub} that this would even not
be sufficient to get a reachable optical depth, since the annihilation channel 
would be $e^+e^-$ in that case. Moreover, other astrophysical constraints on the
annihilation cross section, from x-rays or gamma rays, must then also be taken 
into account, which can be summarized as $\sigv\lesssim 10^{-31}\,{\rm cm^3/s}\,
(\mchi/{\rm MeV})^2$~\cite{2004PhRvL..92j1301B,2006MNRAS.368.1695A}. Such 
constraints strongly limit the possible increase in the optical depth. 
Other ways to increase the electron density can still be advocated, like the 
presence of black holes in the centers of some subhalos. We discuss this
hypothesis in the next section.

\section{Additional impacts from intermediate-mass black holes}
\label{sec:imbhs}
In this section, we complete the study of Galactic dark sources by considering
a putative population of IMBHs.

\begin{figure*}[t!]%[htp]
 \centering
\includegraphics[width=\columnwidth]{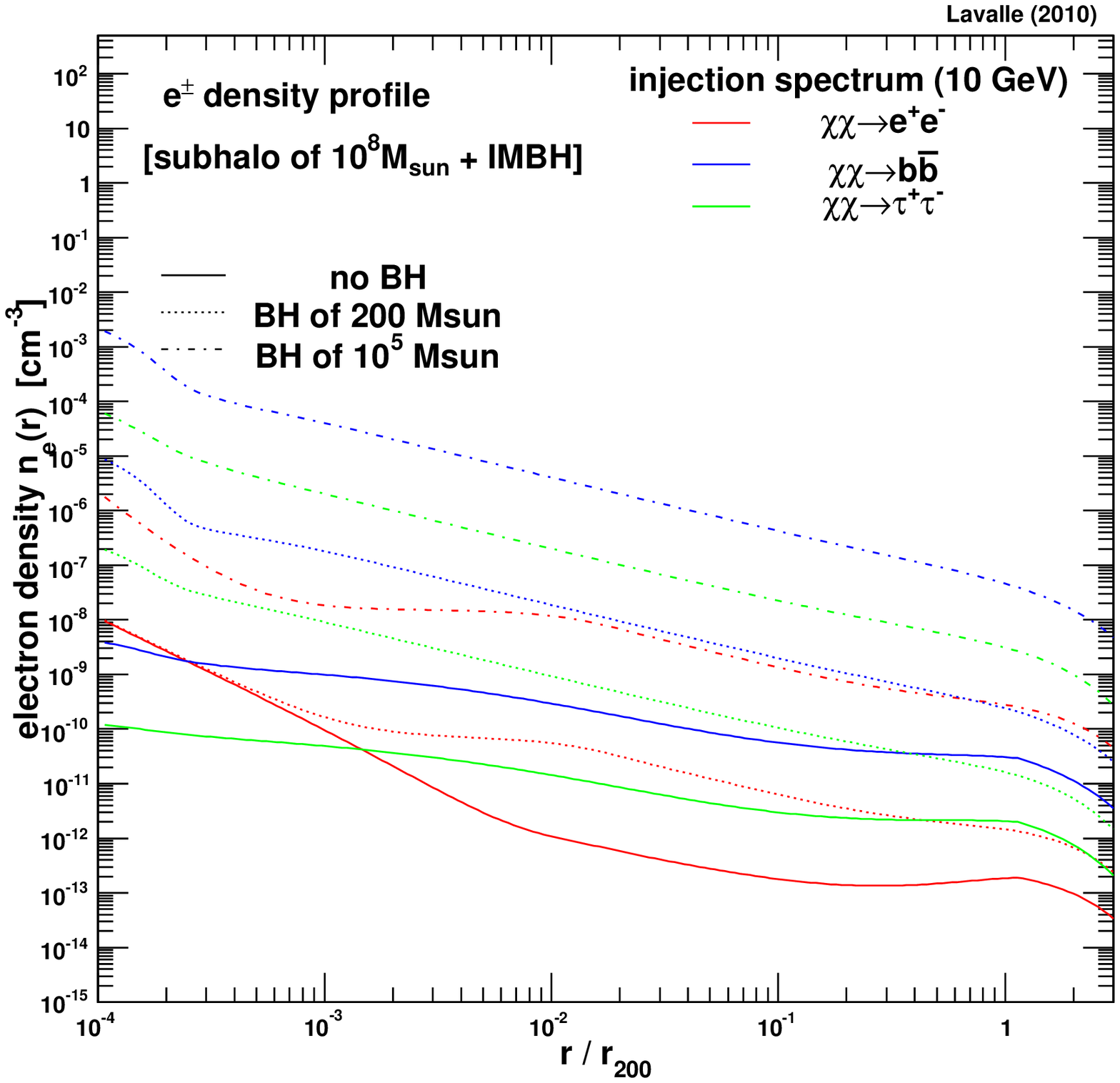}
\includegraphics[width=\columnwidth]{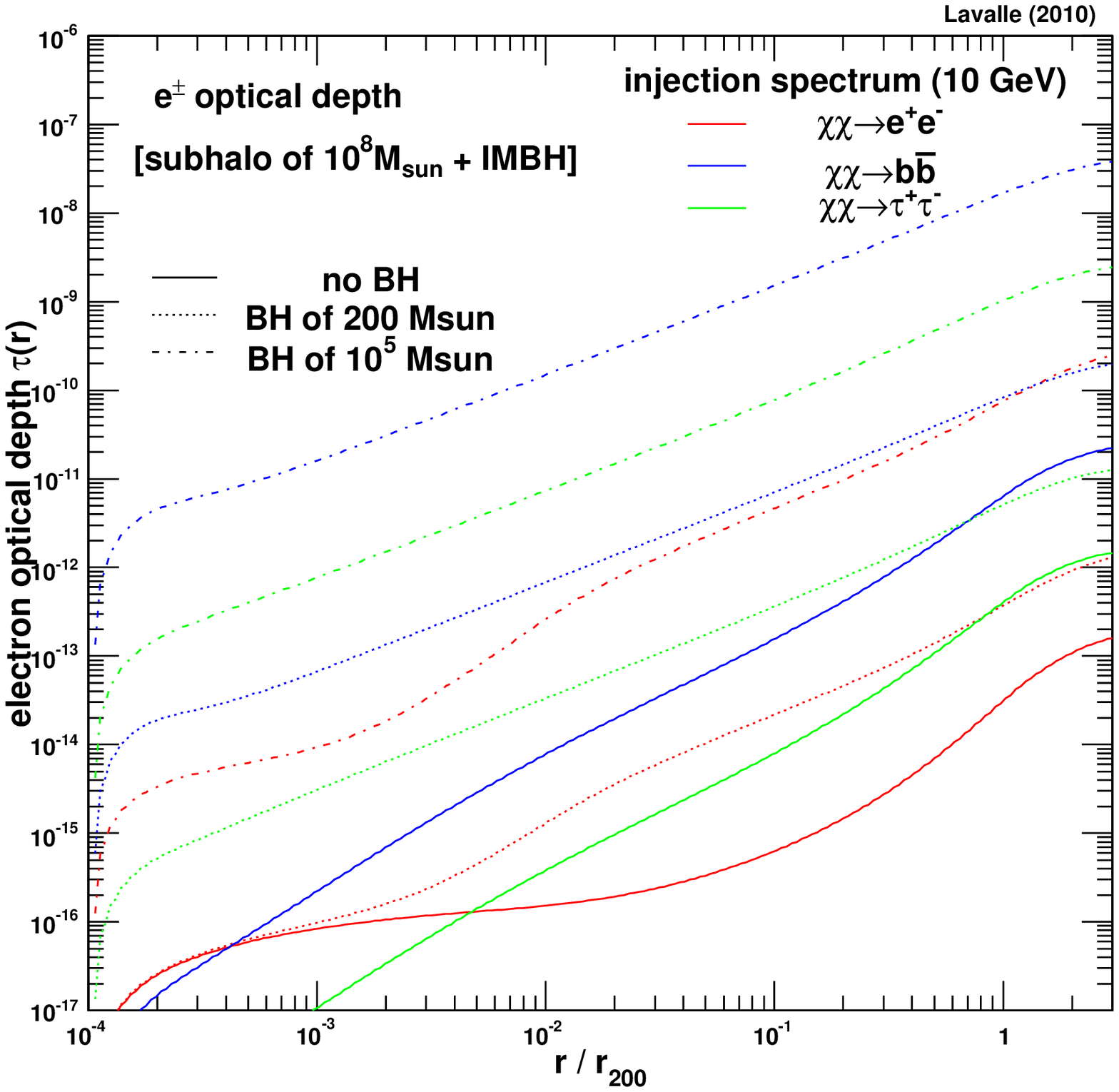}
\caption{Same as \citefig{fig:ne_tot_sub}, except that the solid curves 
  characterize a subhalo without IMBH, the dotted curves the same subhalo 
  hosting a $200\msun$ IMBH, and the dash-dotted curves the same subhalo 
  hosting a $10^5\msun$ IMBH.}
\label{fig:ne_tot_bh}
\end{figure*}

\begin{table}
\centering
\begin{tabular}{cccccc}
\hline
BH mass & $r_h$ & $r_{\rm sp}$ & $\rho_{\rm sp} $ & $r_{\rm sat}$ & $r_{\rm sch}$\\
$[\msun]$ & $[{\rm pc}]$ & $[{\rm pc}]$ & $[10^4\,{\rm GeV/cm^3}]$ & 
$[10^{-3}\,{\rm pc}]$ & $[10^{-11}\,{\rm pc}]$\\
\hline
$200$ & $1.02$  & $0.20$ & $114.81$ & $4.17$ & $1.92$ \\
$10^5$ & $24.41$ & $4.88$ & $4.59$ & $25.17$ & $960$ \\
\hline
\end{tabular}
\caption{Spiky-NFW parameters for a prototype $10^8$ \msun\ subhalo hosting a 
  black hole of (i) 200 and (ii) $10^5$ \msun. The remaining subhalo parameters 
  can be found in~\citetab{tab:cl}.}
\label{tab:bh}
\end{table}

Although the formation scenario of IMBHs is still debated and probably not 
unique, there are many observational hints for their existence, as \eg\ 
ultra-luminous x-ray sources (see \cite{2009Natur.460...73F} for a recent 
case, and \cite{2004IJMPD..13....1M} for a review). Among interesting 
possibilities, IMBHs could be the end products of very massive Population III
stars, those with typical masses $\gtrsim 250\, \msun$ 
\cite{1984ApJ...280..825B}, and thereby seed the supermassive black holes 
observed in most of the galaxies~\cite{2001ApJ...551L..27M}.

If IMBHs are common objects among the first stars, some should still wander in 
the halos of galaxies. One appealing idea is that if they have 
formed from baryon gas cooling in protohalos of dark matter, they could have 
raised minispikes from the adiabatic compression of the surrounding dark matter
\cite{1999PhRvL..83.1719G}, making them excellent Galactic or extragalactic 
targets in the search for dark matter annihilation signals. This idea was 
proposed in Ref.~\cite{2005PhRvL..95a1301Z} for gamma-ray searches, and 
further promoted with more details in Ref.~\cite{2005PhRvD..72j3517B} (see also 
\cite{2008IJMPD..17.1125F} for a recent review). The authors of the latest 
reference discussed scenarios in which the number of Galactic IMBHs --- within a
Galactic radius of $\sim 200$ kpc --- could vary from hundreds to thousands.

The most optimistic scenarios are already in tension with current observations 
in gamma-ray astronomy~\cite{2009PhRvL.103p1301B}, but the general picture is 
still valid and can be probed with new generation large-field-of-view gamma-ray 
instruments, like the Fermi satellite~\cite{2009PhRvD..79d3521T}. For a better 
relevance in the frame of dark matter searches, it is important to detect such 
signals outside the Galactic disk and bulge to escape astrophysical foregrounds
and minimize interpretation issues.
Likewise, it is important to detect complementary signatures which could 
help to confirm or infirm their dark matter origin. Antimatter cosmic rays are 
probably not interesting (i) because the astrophysical background is not under 
control in some cases, (ii) current measurements are compatible with 
astrophysical explanations (see \eg~\cite{2010arXiv1002.1910D} for the Galactic 
electrons and positrons, and \cite{2009PhRvL.102g1301D} for antiprotons) and 
(iii) sources distant by more than a few kpc from the Galactic disk are not 
expected to contribute significantly to the local cosmic-ray flux because of 
the diffusive nature of their propagation in the interstellar medium. As for 
subhalos, we check here whether dark matter annihilation around IMBHs 
could generate any observable SZ signal.

Although some population modelings are available in~\cite{2005PhRvD..72j3517B},
the associated theoretical uncertainties remain to be investigated 
(see~\cite{2010arXiv1008.3552S} for more details on uncertainties). 
Therefore, it seems safer to concentrate the present analysis on individual 
objects without accounting for any putative statistical property. Proceeding so,
we aim at checking whether isolated high-latitude IMBHs could
generate significant SZ contributions in contrast to isolated subhalos.
Nonetheless, before starting, it is interesting to use the averaged properties 
of some minispike scenarios to check their potential as SZ targets. If we
take an average annihilation volume $\langle \xi_{\rm sp} \rangle\sim 10^4 \, 
{\rm kpc^3}$ for minispikes, reminiscent of the most optimistic scenario of
\cite{2005PhRvD..72j3517B} 
(see~\eg~\cite{2007PhRvD..76h3506B,2009PhRvL.103p1301B}), then the total
average annihilation rate in those objects is $\propto {\cal S}_2 \langle 
\xi_{\rm sp} \rangle$. Converting this in terms of an average electron density 
inside a $10^7\msun$ subhalo of scale radius $r_s\sim 100$ pc (annihilation 
to electron-positron pairs), we get the following zeroth order estimate of the 
optical depth:
\ben
\langle \tau^{\chi}_2 \rangle_{\rm sp}
&\sim & 2\,\sigma_T\,{\cal S}_2\,r_s\,\frac{\langle \xi_{\rm sp} \rangle}{V_s} \\
&\sim & 1.95\times 10^{-6}\nn \\ 
& \times & \frac{{\cal S}_2}{1.35\times 10^{-29}\,{\rm cm^{-3}s^{-1}}}\,
\frac{\langle \xi_{\rm sp} \rangle}{10^4\,{\rm kpc^3}} \,
\left[\frac{r_s}{100\,{\rm pc}}\right]^{-2} \;,\nn
\een
where ${\cal S}_2$ was computed using $\mchi = 10$ GeV with the other canonical
values, and where we have assumed that 
$r_0\,{\cal J}_2\sim r_s\,\langle \xi_{\rm sp} \rangle / V_s$. This is thereby 
worth a more detailed investigation.

Aside from statistical properties that may depend on structure formation and 
evolution, considering single IMBHs allows us to motivate a quite generic 
modeling of dark matter distribution around them by simply accounting for the 
adiabatic compression~\cite{1999PhRvL..83.1719G} of the host dark matter 
microhalo during the IMBH growth --- this is also one of the main assumptions 
of Ref.~\cite{2005PhRvD..72j3517B}, upon which the authors plugged an evolution 
history by means of numerical simulations to estimate the survival population 
statistical properties. Thus, starting from a microhalo density profile scaling
like $r^{-\gamma}$, the adiabatic growth of a forming IMBH raises a spike of 
index $\gamma_{\rm sp} = (9-2\gamma)/(4-\gamma)$ by angular momentum 
conservation. For instance, choosing $\gamma=1$ (NFW) implies a spike index of 
$\gamma_{\rm sp} = 7/3\simeq 2$. Further adopting an NFW initial profile as a 
generic case, the final dark matter density shape around the IMBH can be 
described as
\begin{align}
\rho(r) = 
\begin{cases}
\rho_{\rm sat} \;\;\;\; 
& r_{\rm sch} < r \leq r_{\rm sat}\\
\rho_{\rm sp} \left( \frac{r}{r_{\rm sp}} \right)^{-\gamma_{\rm sp}} 
& r_{\rm sat} < r \leq r_{\rm sp} \\
\rho_s\frac{(r/r_s)^{-1}}{(1+r/r_s)^2} & r > r_{\rm sp} 
\end{cases}
\label{eq:rho_imbh}
\end{align}
where the subscript {\em sp} is related to the spike (density, extent, index), 
$r_{\rm sat}$ is the radius at which the annihilation rate saturates~
\cite{1992PhLB..294..221B}, and $r_{\rm sch}=2\,G\,m_{\rm bh}/c^2\simeq 9.6\times
10^{-14}{\rm pc} \,(m_{\rm bh}/\msun) $ is the Schwarzschild radius of a 
black hole of mass $m_{\rm bh}$ below which neither particles nor light can get 
out~\cite{1916AbhKP......189S}. We have the implicit relation $\rho_{\rm sp} 
\simeq \rho_s \, r_s/r_{\rm sp} $, provided $r_{\rm sp}\ll r_s$. The actual 
spatial scales can be inferred from the radius of gravitational influence 
$r_h$ of the black hole. It was indeed found in Ref.~\cite{2005PhRvD..72j3517B} 
that $r_{\rm sp} \approx 0.2 \,r_h$ in most cases. Furthermore, it turns out 
that $r_h$ can be related to $r_s$ from the implicit equation
~\cite{2004cbhg.symp..263M,2006RPPh...69.2513M},
\ben
M(<r_h) = 4\,\pi \int_0^{r_h}dr\,r^2\,\rho(r) = 2\, m_{\rm bh}\;,
\label{eq:m_rh}
\een
which is analytical in the NFW case:
\ben
\label{eq:def_rh}
\frac{2\, m_{\rm bh}}{4\,\pi} &=& 
\rho_s \, r_s^3 \left[ \ln\left\{ \frac{r_h+r_s}{r_s} \right\}
-\frac{r_h}{r_h+r_s}\right] \\
&\approx & \frac{\rho_s\, r_s\, r_h^2}{2}\;.
\label{eq:approx_rh}
\een
The last line was obtained with the limit $r_h\ll r_s$, eventually leading to
\ben
\label{eq:r_bh}
r_h &\approx&  \sqrt{\frac{\,m_{\rm bh}}{\pi\,\rho_s\,r_s}}  \\
&\approx& 3.9\,{\rm pc} \,
\left[ \frac{m_{\rm bh}}{100\,\msun} 
\left[ \frac{\rho_s}{10\,{\rm GeV/cm^3}} \frac{r_s}{100\,{\rm pc}} \right]^{-1} 
\right]^{1/2} \nn \;.
\een
Note that this approximation is only valid for $r_h\ll r_s$; \citeeq{eq:def_rh} 
must be used otherwise. Now, although the spike radius should in principle 
be computed numerically, we can use the scaling relation $r_{\rm sp} \approx 0.2
\,r_h$~\cite{2004cbhg.symp..263M}. Finally, all these spatial scales have to be 
compared with the saturation radius defined by the saturation 
density~\cite{1992PhLB..294..221B},
\ben
\label{eq:rho_sat}
\rho_{\rm sat} \equiv \rho(r_{\rm sat}) &\approx& 
\frac{ \mchi }{\sigv t_{\rm bh} } \\
&\approx & 10^8\,{\rm GeV/cm^3} \nn\\
&& \times\, \frac{\mchi}{\rm GeV} \left[ 
\frac{\sigv }{3\times 10^{-26}{\rm cm^3/s}} 
\frac{t_{\rm bh}}{10\,{\rm Gyr}}\right]^{-1} \;,\nn
\een
where $t_{\rm bh}$ is the black-hole age.

We now take a template example relying on the study of single subhalos we 
performed in~\citesec{subsubsec:diff}, implying quite a massive subhalo of 
$10^8\msun$, the NFW profile of which has now to be compressed and develop a 
spike because of the presence of an IMBH in its center. We consider two 
IMBH mass cases reminiscent of the scenarios proposed 
in~\cite{2005PhRvD..72j3517B}, a {\em soft} case with $m_{\rm bh} = 200$ 
\msun\ and a {\em strong} case with $m_{\rm bh} = 10^5$ \msun (see also
\cite{2008IJMPD..17.1125F}). The associated parameters that we derived according
to Eqs.~(\ref{eq:m_rh}-\ref{eq:rho_sat}) are listed in~\citetab{tab:bh}. We 
are thus armed to calculate the electron density arising from dark matter 
annihilation in such objects, using the same diffusion-loss transport model as 
in~\citesec{subsubsec:diff}.

We plot our results in~\citefig{fig:ne_tot_bh}, where the electron density 
profile is reported in the left panel and the corresponding cumulative
optical depth appears in the right panel. We compare three different 
configurations: a single $10^8$ \msun~subhalo (solid curves, same as 
in~\citefig{fig:ne_tot_sub}), the same subhalo with a $200$ \msun~central
black hole (dotted curves), and with a $10^5$ \msun~central black hole 
(dash-dotted curves). As before, we assumed a 10 GeV WIMP with the 
annihilation channels discussed above. As expected, we see that the presence
of a central black hole (through its related spike) has drastic consequences
on predictions. The optical depth is shown to increase by 1 (3) order of 
magnitude provided a spike raised by a $200$~($10^5$) \msun~central black hole,
independently of the injection spectrum. This can actually be related to 
the increase of the annihilation rate averaged over the diffusion length, 
which, unfortunately, has no analytical form. The optical depth can reach 
$\tau_e\sim 10^{-7}$ in the most favorable $b\bar{b}$ spectral case, but such a 
level is still, unfortunately, irrelevant to observation.

\section{Conclusion}
\label{sec:concl}

In this paper, we have studied the SZ effect potentially generated by dark 
matter annihilation (or decay) products on the Galactic scale, in the 
high-latitude sky. We have focused our analysis on (i) the smooth Galactic halo 
(see~\citesec{subsubsec:smooth}), (ii) subhalos (see~\citesec{subsubsec:sub}) 
and (iii) putative spikes raised by black holes in the centers of individual 
subhalos (see~\citesec{sec:imbhs}). We have considered canonical properties for 
annihilating (or decaying) dark matter, though in the very light mass range 
below 10 GeV.

For the smooth-halo contribution, we have derived our predictions in the 
diffusionless limit of the electron transport, which is valid
whenever the electron injection rate does not vary significantly over a 
diffusion length. In this approximation, we have found that 
$\tau_e\lesssim 10^{-8}$ for a 1 GeV annihilating or decaying WIMP, rescaled
to $\lesssim 10^{-10}$ for a 10 GeV annihilating WIMP. The average contribution
of subhalos was then shown to boost the signal by 1 order of magnitude
at most in the case of dark matter annihilation, but only when using a rather 
favorable subhalo phase space (mass index of 2) --- the effect is strongly 
diminished for decaying dark matter, since the injection rate then scales like 
the density, not like the squared density; we did not further explore the 
impact of subhalos in this context.

The study that we performed on isolated subhalos has led us to abandon the 
diffusionless limit of electron transport, which was shown unjustified 
for a source exhibiting a strong spatial gradient over the typical diffusion 
length (like the central cusps of galaxy clusters~\cite{2010JCAP...02..005L}). 
Indeed, a $10^8$ \msun~subhalo, \ie~quite massive, has a radial extent of a few 
kpc, of the same order as the diffusion length. We have notably illustrated how 
a core radius emerges in the electron distribution because of spatial diffusion,
independently of injection spectra. This smearing strongly dilutes the electron 
density so that the optical depth cannot reach values of interest. For a quite 
massive subhalo of $10^8$ \msun, we found an optical depth 
$\tau_e \lesssim 10^{-11}$ for a 10 GeV annihilating WIMP with canonical 
properties. These results are in agreement with those obtained in similar 
studies on dwarf spheroidal galaxies~\cite{2007PhRvD..75b3513C,2010arXiv1005.2325H}.

Finally, we checked whether dark matter spikes raised by IMBHs in massive 
subhalos from adiabatic compression could lead to observable SZ signals. To 
proceed, we have designed a generic spike modeling featured by the properties 
of the host subhalo and by the IMBH mass. We have focused on a template example
consisting in a 200~($10^5$) \msun~central black hole located at the center of 
a $10^8$ \msun~subhalo. We have shown that such a spike could boost the optical
depth by 1  (3, respectively) order of magnitude, which is in fact related
to the increase in the annihilation rate averaged over a typical diffusion 
length. Nevertheless, even such an increase is not enough to get an electron
density large enough for leaving a SZ imprint in the CMB sky. We found $\tau_e 
\lesssim 10^{-7}$ for our 10 GeV WIMP, which leads us to conclude that dark 
matter is globally not expected to generate SZ fluctuations on the Galactic 
scale. Because of these quite modest values obtained for the optical depth, it 
is not necessary to go deeper in the spectral analysis to derive the full SZ 
distortion spectrum~\cite{2010JCAP...02..005L}.

Note that when dealing with isolated massive subhalos, hosting IMBHs or not,
we have computed the electron densities in the frame of a diffusion-loss
transport model for which we assumed a local diffusion coefficient and 
an energy loss driven by interactions with CMB only. Though the latter
hypothesis seems reasonable far away from the Galactic disk in a medium 
almost devoid of gas and stars, the former is more difficult to justify, since 
electrons should be close to free propagation in such regions. Nevertheless, we 
have argued that considering a more realistic transport would actually 
strengthen the advocated smearing effect coming from spatial diffusion, since 
the diffusion coefficient should then be increased, which in fact makes our 
predictions rather conservative. Still, more accurately relating the 
phenomenology of transport to the interstellar medium properties remains a
vast topic to undertake so as to improve all analyses focused on nonthermal 
cosmic-ray electron-induced radiations.

\acknowledgments
We are indebted to C\'eline B{\oe}hm for stimulating discussions and for earlier
collaborations on related topics.

% \cite{2000A&A...360..417E} Ensslin
% \cite{1984ApJ...280..825B} Bond et al
% \cite{2004IJMPD..13....1M} Miller & Colbert
% \cite{2001ApJ...551L..27M} Madau & Rees
% \cite{2001ApJ...554..548F} Fryer & Kalogera
\bibliography{lavalle_bib}

%%%%%%%%%%%%%%%%%%%%%%%%%%%%%%%%%%%%%%%%%%%%%%%%%%%%%%%%%%%%%%%
\end{document}